\documentclass[11pt,a4paper]{article}
\usepackage{graphicx,amsmath,parskip,subcaption,enumerate}
\usepackage{float}
\usepackage{caption}
\usepackage[margin=1in]{geometry}
\usepackage[hidelinks]{hyperref}
 \usepackage[sorting=none, backend=bibtex]{biblatex}
\renewbibmacro{in:}{}
\DefineBibliographyStrings{english}{%
  page             = {},
  pages            = {},
} 
\DeclareDelimFormat{finalnamedelim}{\addcomma\space}
\usepackage{booktabs}
\usepackage{xcolor}
\allowdisplaybreaks

\begin{document}
\begin{titlepage}



\vspace{10mm}
\begin{center}
{\Large \bf Closed timelike curves in modified theories of  gravity }

\vskip 1. cm
  {
   Agnik Sen Roy$^{a}$\footnote{e-mail : agniksr@gmail.com}, 
  Sashideep Gutti$^{b}$\footnote{e-mail : sashideep@hyderabad.bits-pilani.ac.in},  
  Shailesh Kulkarni$^{a}$\footnote{e-mail : shailesh.kulkarni@unipune.ac.in} 
  }

\vskip 0.5cm

{\it $^a\,$ Department of Physics, Savitribai Phule Pune University, Ganeshkhind, Pune, 411007, India}\\
{\it $^b\,$ Birla Institute of Technology and Science, Pilani, Hyderabad Campus, Hyderabad, 500 078, India}\\
\end{center}

\thispagestyle{empty}

\vskip1.5cm


\centerline{\bf ABSTRACT} \vskip 4mm

\vspace{1cm}

\noindent We study closed timelike curves (CTCs) in three rotating spacetimes arising in modified
gravity: the tidally charged braneworld black hole, the Kerr-Newman black hole in Rastall
gravity, and the charged Johannsen-Psaltis spacetime. Using the sign of $g_{\phi\phi}$ on the
equatorial plane, we show that an additional parameter of each geometry controls the
chronology-violating region, and we obtain critical parameter values analytically. In the
braneworld case, the positive $r$ CTC region exists only for positive tidal charge and
disappears completely for $q\leq 0$. In Rastall gravity, the extent of the CTC region depends jointly on the Rastall coupling and the surrounding matter. For each matter type we find a finite range of the coupling
for which the CTC region is absent.
In the charged Johannsen-Psaltis spacetime, the deformation parameter can either increase or
decrease $g_{\phi\phi}$ depending on the radial position, and we derive a window of its values
for which the CTC region is removed from the radial domain considered. Thus, in all three examples, the additional parameters can be chosen to control the extent of
equatorial CTCs, and even to eliminate them entirely.
\vspace{2cm}

\end{titlepage}
\setcounter{footnote}{0}

\section{Introduction}

One of the important features of general relativity is that the geometry
of spacetime determines its causal structure. In the majority of physically meaningful spacetimes, the light-cone structure serves to distinguish between events that occurred in the past and those that will occur in the future. However, Einstein's field equations also admits for the solutions that contain closed timelike curves (CTCs). The temporal ordering of events is lost along CTCs, raising questions concerning causality
and predictability in general relativity. Regions of spacetime in causal contact with
CTC-containing regions are also problematic from the viewpoint of numerical relativity, since
they do not admit a well-posed initial-value formulation.
 The solutions of Einstein's field equations admitting CTCs were first discussed in \cite{Godel:1949ga}. The causal structure of the Kerr family was analysed extensively by Carter \cite{Carter:1968rr}. Other notable examples of spacetimes admitting closed timelike curves include the Gott spacetime \cite{Gott:1990zr} and wormhole geometries \cite{Visser:1995cc}. For general discussions of causal structure and CTCs, see \cite{Hawking:1973uf, Frolov:1998wf}.
The existence of CTCs in mathematically consistent solutions raises a significant question as to whether it  is possible for such causality violating regions to exist in physically relevant spacetimes, or is there a physical mechanism that prevents their formation. This question is closely related to Hawking's chronology protection conjecture
\cite{Hawking:1991nk}, according to which the physical effects associated with the formation of a
chronology horizon may prevent CTCs from appearing. This possibility has also been studied 
from the perspective of quantum field theory in curved spacetime
\cite{Klinkhammer:1992tb, Kay:1996hj}. However, a general proof of the mechanism remains
unknown.

The Kerr-Newman (KN) spacetime is a prototype example in which the problem of CTCs can be examined more closely. In the KN spacetime, the CTC region lies between the inner horizon and the naked singularity, so that the causal connectivity between the two makes the evolution of Cauchy data an ill-posed problem \cite{Hawking:1973uf}. Furthermore, this causality violation cannot be removed by passing to a suitable covering space \cite{Carter:1968rr}. For stationary axisymmetric spacetimes, including the Kerr family, the Killing vector field $\xi_{(\phi)}=\partial_\phi$ associated with rotations has norm $\xi^2_{(\phi)}=g_{\phi\phi}$. Its closed integral curves become CTCs when $g_{\phi\phi}<0$, which therefore provides the criterion for identifying the CTC region. We then ask whether modifications of the gravitational theory can alter this region. The additional parameters appearing in modified gravity theories and their black-hole solutions may give a handle on the extent of the CTC region and, for suitable values, may reduce or eliminate it

There are many extensions or modifications of the KN geometry that contain
additional parameters apart from the usual mass, rotation and electric
charge. The origin of these parameters depends on the particular model.
They may arise, for example, from extra-dimensional effects, modified
matter-curvature coupling, or a parametrised deformation of the Kerr
geometry. Since these parameters modify the metric coefficients, they
can also modify $g_{\phi\phi}$ and hence the location and extent of the
CTC region. This possibility was considered in our earlier work
\cite{Prasad:2018hdj}, where the chronology protection
problem was studied for two modified KN geometries: the
noncommutative geometry inspired KN black hole \cite{Modesto:2010rv} and the
KN black hole in $f(R)$ gravity \cite{Cembranos:2011sr}.
In both cases, the additional parameters were found to modify the
chronology-violating region. For the noncommutative geometry inspired
solution, an appropriate range of the noncommutative parameter removed
the CTC region together with the Cauchy horizon. In the $f(R)$ case,
the CTC region could also be removed mathematically, but the required
parameter values were not compatible with the relevant observational
constraints. These results showed that modifications of the
KN geometry can have an important effect on its causal
structure.

Another question is whether the existence or removal of CTCs is 
related to the energy conditions. Hawking's analysis \cite{Hawking:1991nk} already showed the importance of the averaged weak
energy condition in the formation of chronology-violating regions. 
The relation between causality violation and energy conditions also
appears naturally in the study of traversable wormholes and time-machine
geometries. In particular, the role of exotic matter and the violation
of the standard energy conditions in such spacetimes is discussed
extensively in
\cite{Visser:1995cc}. The connection between CTCs and energy conditions in regular
spacetimes was studied more directly in our subsequent work
\cite{Gutti:2022qsv}. In that work, the possible
role of violations of energy conditions in the elimination of CTCs
was investigated. More recently, the chronology protection conjecture in a rotating black-hole solution of Lorentz-violating gravity has been studied in \cite{ParkMaeda2026}.
In their solution the causality-violating region remains the same as that
of the corresponding KN or KN-(A)dS spacetime.
Chronology protection is instead realised through the appearance of an
additional curvature singularity at the boundary of this region, which
prevents exterior observers from accessing the chronology-violating
domain. 

In the present work, we investigate 
the effect of additional parameters appearing in  modified geometries on the  location and extent of the
chronology violating region.  We consider the tidally charged braneworld black hole, the
KN black hole in Rastall gravity, and the charged
Johannsen-Psaltis spacetime.
The braneworld framework and its gravitational consequences have been
reviewed in
\cite{MaartensKoyama2010}, while the rotating braneworld solution
used here was obtained in
\cite{Aliev:2005bi}.
 Rastall gravity was proposed
in~\cite{Rastall1976}, and black holes surrounded by perfect fluid in this
framework were studied in~\cite{Heydarzade:2017wxu}. 
The parametrised rotating geometry used in the third example was
introduced in \cite{JohannsenPsaltis2011}, and its charged extension was obtained in \cite{RahimSaifullah2019}.
These three geometries have different physical origins. The braneworld
solution contains a tidal charge arising from higher-dimensional
gravitational effects. In Rastall gravity, the geometry depends on the
modified matter--curvature coupling and on the surrounding matter
distribution. The charged Johannsen--Psaltis spacetime contains a
deformation parameter describing a departure from the standard
KN geometry. These three models  provide three different settings
in which the effect of additional parameters on chronology violation
can be examined. For each of these models, 
we consider the positive-$r$ equatorial
CTC region and the local causal behaviour at fixed radial positions. We determine how variation of the relevant parameter changes
the boundary of the CTC region and identify parameter ranges for which
the CTC region is reduced or eliminated.  

The paper is organised as follows. In Sec.~\ref{sec:Tidal} we consider the tidally charged braneworld black hole and study the dependence of the CTC region on the tidal charge. Sec.~\ref{sec:Rastall} discusses the KN black hole in Rastall gravity for dust, dark energy and perfect dark matter
backgrounds, and examine the role of the Rastall coupling. In Sec.~\ref{sec:JPKN} we study the charged Johannsen--Psaltis spacetime and analyse the effect of the deformation parameter on the chronology-violating region. Finally, in Sec.~\ref{sec:Conclusion} we summarise the main results and discuss possible extensions of this work.

\section{Models}

\subsection{Tidally Charged Braneworld Black Hole}\label{sec:Tidal}

The braneworld scenario provides one of the simplest extensions of general relativity wherein our observable Universe is regarded as a $(3+1)$-dimensional hypersurface (brane) embedded in a higher-dimensional bulk spacetime. Such extra-dimensional scenarios were developed, in
particular, in the Randall-Sundrum models
\cite{Randall:1999ee,Randall:1999vf}. While the Standard Model fields are limited to the brane, gravitational degrees of freedom extend into the extra dimensions. This leads to corrections to the effective gravitational field equations on the brane due to the higher-dimensional geometry \cite{Shiromizu:1999wj}. For a detailed review of braneworld gravity, see \cite{MaartensKoyama2010}.

A consequence of these higher-dimensional corrections is the appearance of the projected bulk Weyl tensor in the effective four-dimensional field equations. Even in vacuum, these non-local gravitational effects modify the exterior geometry of a black hole through an additional parameter known as the \emph{tidal charge} \cite{Dadhich:2000am}. Unlike the electric charge of the KN solution, the tidal charge has a purely gravitational origin and can allow for positive or negative values. This feature distinguishes braneworld black holes from their counterparts in general relativity and leads to the modifications of the horizon structure and other strong-field properties.

For our purpose, the braneworld black hole offers an ideal framework for analyzing the effect of extra dimensions on CTCs. Since the tidal charge directly modifies the spacetime geometry, it is reasonable to expect that it may also alter the region where chronology violation occurs. If suitable values of the tidal charge eliminate the CTC region, this would indicate that higher-dimensional gravitational effects can restore the chronology at least locally without any additional matter fields.

The rotating charged braneworld black hole was obtained in \cite{Aliev:2005bi} by adopting a Kerr ansatz for the induced metric on the brane and solving the corresponding effective gravitational field equations. The resulting metric is  similar to the KN geometry after replacing the square of the electric charge by the tidal charge parameter,
$
Q^2 \rightarrow q,
$
where the parameter $q$ is allowed to be either positive or negative. The metric is given by

\begin{align}
ds^2 =\;&
-\left(1-\frac{2Mr-q}{\rho^2}\right)dt^2
-\frac{2a(2Mr-q)\sin^2\theta}{\rho^2}\,dt\,d\phi
+\frac{\rho^2}{\Delta(r)}\,dr^2
+\rho^2\,d\theta^2 \nonumber\\[1mm]
&
 +\sin^2\theta
\left(
r^2+a^2
+\frac{a^2(2Mr-q)\sin^2\theta}{\rho^2}
\right)d\phi^2
\label{eq:braneworld_metric}
\end{align}

with

\[
\Delta(r)=r^2-2Mr+a^2+q,
\qquad
\rho^2=r^2+a^2\cos^2\theta
\]

We now restrict our attention
to $g_{\phi\phi}$ on the equatorial plane
($\theta=\pi/2$). Then $\rho^2=r^2$, and 
\begin{equation}
g_{\phi\phi}
=
\frac{\mathcal{P}(r)}{r^2},
\qquad
\mathcal{P}(r)
=
r^4+a^2r^2+2Ma^2r-a^2q \label{brane-gphiphiequator}
\end{equation}

The behaviour of this function determines the presence or absence of
CTCs in the spacetime.
For \(r>0\), the sign of
\(g_{\phi\phi}\) is completely determined by
\(\mathcal{P}(r)\). The (radial) boundary of the CTC region is thus
obtained by setting $\mathcal{P}(r)=0$. 
The sign of the tidal charge has an immediate effect on
the causal structure. For \(q<0\), all the terms in
\(\mathcal{P}(r)\) are positive for \(M>0\) and \(r>0\).
Therefore, $g_{\phi\phi}$ remains positive for all
$r>0$ whenever $q<0$ and thus no equatorial CTC region is present. The same conclusion
holds for \(q=0\). 
Therefore, the Kerr limit does not contain an equatorial CTC
region at positive values of the Boyer--Lindquist radial
coordinate.

For \(q>0\), on the other hand, the tidal-charge term
dominates near the origin. In this case,
\begin{equation}
\lim_{r\rightarrow0^+}g_{\phi\phi}
=
-\infty, \ \ \lim_{r\rightarrow\infty}g_{\phi\phi}
=
+\infty.
\end{equation} 

Moreover, for $M>0$, $a\neq 0$, and $r>0$, $$\frac{d\mathcal{P}}{dr} = 4r^3+2a^2r+2Ma^2>0$$

Thus, $\mathcal{P}(r)$ is a strictly increasing function on the
positive radial axis. Also, 
\begin{equation}
\mathcal{P}(0)=-a^2q<0,
\qquad
\lim_{r\rightarrow\infty}\mathcal{P}(r)=+\infty.
\end{equation}
It then follows that $\mathcal{P}(r)$ has exactly one positive root,
which we denote by $r_{\rm CTC}$. Therefore,
$g_{\phi\phi}<0$
for $0<r<r_{\rm CTC}
$, 
whereas $g_{\phi\phi}>0$ for $r>r_{\rm CTC}$. 

\begin{figure}[H]
    \centering
     \includegraphics[width=0.5\linewidth]{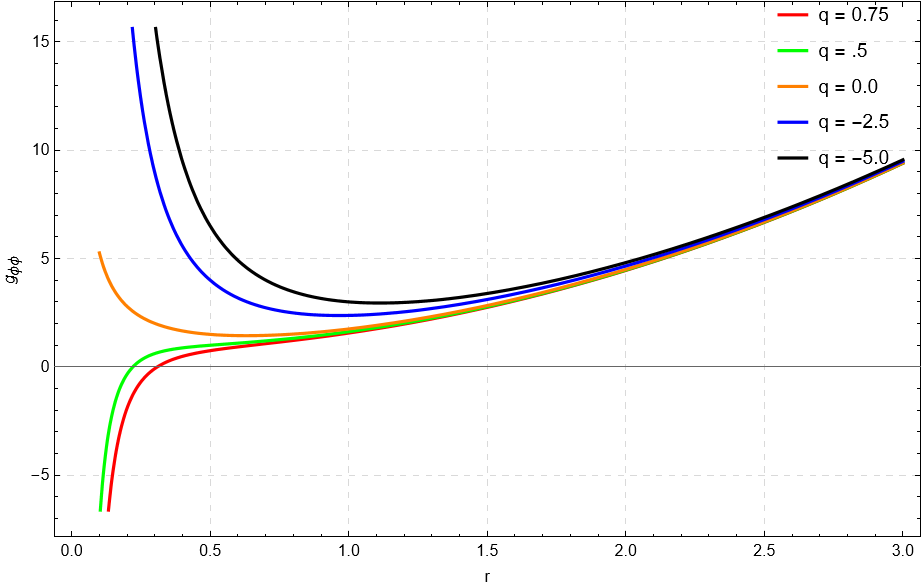}
   \caption{
The equatorial azimuthal metric component
$g_{\phi\phi}/M^2$
as a function of the dimensionless radial coordinate
$r/M$ for fixed $a/M=0.5$ and different values of
$q/M^2$ satisfying the bound $M^2-a^2-q \ge 0$. The region satisfying
$g_{\phi\phi}<0$ corresponds to axial closed timelike
curves. Positive tidal charge produces a chronology-violating region at small positive $r$, whereas
$q\le0$ removes this region from the positive-$r$
equatorial domain.
}
    \label{fig:brane-gphiphi-r}
\end{figure}

At this point, it is useful to locate the CTC region relative to the horizon structure for arbitrary $q>0$. For
$M^2-a^2-q\geq0$, the inner horizon is located at 
 $r_-=M-\sqrt{M^2-a^2-q}$. Expressing the charge $q$ in terms of $M, a$ and $r_{-}$ and substituting it into  Eq.~(\ref{brane-gphiphiequator}), we find 
\begin{equation}
\mathcal{P}(r_-) = r_-^4+a^2r_-^2+2Ma^2r_--a^2q = \left(r_-^2+a^2\right)^2 > 0,
\end{equation}
which is independent of the charge $q$. Since $\mathcal{P}(r)$ is a strictly increasing function for $r>0$ and
$\mathcal{P}(r_{\mathrm{CTC}})=0$, it follows that $r_{\mathrm{CTC}}<r_-$ for every $q$ in
the range $0<q\leq M^2-a^2$. Thus, whenever the positive-$r$
equatorial CTC region is present, it lies entirely inside the inner horizon and is therefore causally disconnected from exterior observers in
the same manner as in the KN black holes. 
Fig.~\ref{fig:brane-gphiphi-r} shows
the radial behaviour of $g_{\phi\phi}$ for different values of
the tidal charge. The curves corresponding to positive $q$ enter
the region $g_{\phi\phi}<0$, while those with $q\leq0$ remain
positive throughout the positive radial domain.

In order to understand how the tidal charge controls the
causal character of a fixed spacetime point, we now keep
the radius fixed at \(r=r_0\) and vary \(q\). At this fixed
radius,
\begin{align}
g_{\phi\phi}(q;r_0)
={}&
r_0^2+a^2+\frac{2Ma^2}{r_0}
-\frac{a^2}{r_0^2}q .
\end{align}
Thus \(g_{\phi\phi}\) depends linearly on the tidal charge,
with
\begin{equation}
\frac{\partial g_{\phi\phi}}{\partial q}
=
-\frac{a^2}{r_0^2}<0.
\label{threecases}
\end{equation}
The critical value at which the chosen radial point lies on
the boundary of the CTC region is determined  by the condition $g_{\phi \phi}(q_\mathrm{crit},r_0)=0$, which gives
\begin{equation}
q_{\mathrm{crit}}(r_0)
=
2Mr_0
+
\frac{r_0^2(r_0^2+a^2)}{a^2}.
\end{equation}
Therefore,
\begin{equation}
\begin{aligned}
q<q_{\mathrm{crit}}(r_0)
&\quad\Longrightarrow\quad
g_{\phi\phi}(r_0)>0,\\
q=q_{\mathrm{crit}}(r_0)
&\quad\Longrightarrow\quad
g_{\phi\phi}(r_0)=0,\\
q>q_{\mathrm{crit}}(r_0)
&\quad\Longrightarrow\quad
g_{\phi\phi}(r_0)<0.
\end{aligned}
\end{equation}

Since $g_{\phi\phi}(r_0,q)$ is strictly decreasing function of  $q$ (see Eq.~(\ref{threecases})), the three cases
above are not merely possibilities to be checked numerically: any
$q>q_{\mathrm{crit}}(r_0)$ automatically places $r_0$ inside the CTC region, and any
$q<q_{\mathrm{crit}}(r_0)$  places it outside. This guarantees, for any chosen
$r_0$, the existence of a parameter range in which the point starts inside the CTC region and
can be moved out of it by decreasing $q$ through $q_{\mathrm{crit}}(r_0)$.

\begin{figure}[H]
    \centering
    \includegraphics[width=0.5\linewidth]{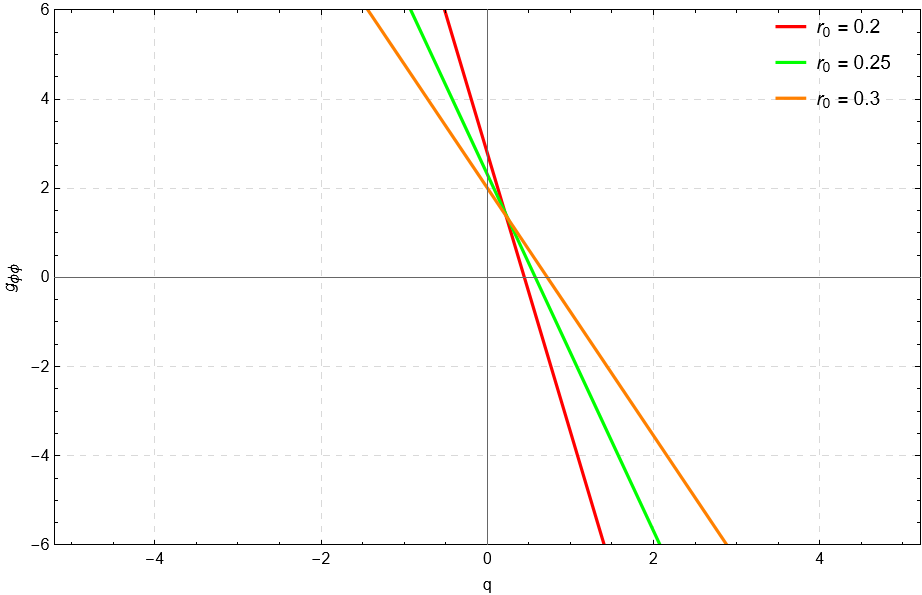}
    \caption{
The azimuthal metric component
$g_{\phi\phi}$ as a function of the tidal charge
$q$ at a fixed radius
$r=r_0$, with $M$ and $a$ held fixed at $1$ and $0.5$ respectively.
}
    \label{fig:brane-gphiphi-q}
\end{figure}
Fig.~\ref{fig:brane-gphiphi-q} illustrates this point more clearly. The horizontal line
$g_{\phi\phi}=0$
separates the causal and chronology-violating regions.
The curve crosses this line at the critical value
$q=q_{\mathrm{crit}}(r_0)$.
Decreasing the tidal charge through this value changes
$g_{\phi\phi}$ from negative to positive, thereby
removing the chosen spacetime point from the CTC region. It is important to distinguish this local transition from the
elimination of the complete CTC region. For
$q<q_{\mathrm{crit}}(r_0)$, the particular point
$r=r_0$ lies outside the chronology-violating region,
however CTCs may still exist at smaller
radii if $q>0$ (i.e., when $0<q<q_{\mathrm{crit}}(r_0)$). The complete positive-$r$ equatorial CTC
region disappears only when $q\le0$. 
This
shows that the tidal-charge
parameter can eliminate the CTCs locally.

The tidal charge therefore acts as a {\it control} parameter for
chronology violation. A positive tidal charge produces a
KN-like CTC region at positive values of the
radial coordinate, whereas decreasing the tidal charge
progressively reduces its extent. At $q=0$, the
positive-$r$ equatorial CTC region disappears, and for
$q<0$, the extra-dimensional contribution ensures that
$g_{\phi\phi}$ remains positive throughout the positive
radial domain. In this restricted sense, non-positive
tidal charge provides a local chronology-protection
mechanism within the braneworld geometry.


\subsection{Kerr-Newman Black Hole in Rastall Gravity}\label{sec:Rastall}
Rastall gravity, proposed in  \cite{Rastall1976} is a 
modification of general relativity in which the usual conservation law
for the energy-momentum tensor is relaxed. Instead of
$\nabla_\mu T^{\mu\nu}=0$, one assumes

\begin{equation}
\nabla_\mu T^{\mu\nu}
=
\lambda\nabla^\nu R,
\end{equation}

where $R$ is the Ricci scalar and $\lambda$ is the Rastall parameter.
In regions where the Ricci scalar is constant, the standard
conservation law is recovered. The corresponding gravitational field
equations are

\begin{equation}
G_{\mu\nu}
+
\kappa\lambda g_{\mu\nu}R
=
\kappa T_{\mu\nu},
\end{equation}

where $\kappa$ denotes the gravitational coupling constant. The
dimensionless combination $\kappa\lambda$ characterises the deviation
from general relativity, while the limit
$\kappa\lambda=0$ reproduces Einstein gravity.

Rastall gravity has attracted considerable attention because it
provides a simple phenomenological extension of general relativity
while remaining consistent with a variety of astrophysical and
cosmological observations; see, for example, \cite{Batista:2011nu}. The  neutron-star models constructed within
Rastall gravity closely reproduce those of general relativity \cite{Oliveira2015}, while the black-hole solutions surrounded by
perfect-fluid matter have been studied in \cite{Xu2017}. These results make Rastall gravity a useful
candidate for investigating the influence of the additional
matter-curvature coupling term on strong gravitational fields.

Heydarzade and Darabi \cite{Heydarzade:2017wxu} derived the complete class of spherically symmetric black hole solutions surrounded by a perfect fluid in Rastall theory, using the equation of state $p = \omega_s\rho$ to parametrise different matter environments. The rotating charged KN black hole surrounded by a perfect
fluid in Rastall gravity was obtained by applying the Newman--Janis
algorithm to the corresponding static solution
\cite{Xu2017}. The resulting spacetime is described by the
metric

\begin{eqnarray}
    ds^2 = &-&\left( 1-\frac{2Mr}{\rho^2} + \frac{q^2}{\rho^2} - \frac{N_sr^{\mu}}{\rho^2} \right)\ dt^2 - \frac{2a(2Mr-q^2+N_sr^{\mu})\sin^2\theta}{\rho^2}\ dtd\phi \nonumber\nonumber\\ 
    &+& \frac{\rho^2}{\Delta(r)}\ dr^2
    + \rho^2\ d\theta^2 
    + \sin^2\theta \left( r^2+a^2+\frac{a^2(2Mr-q^2+N_sr^{\mu})\sin^2\theta}{\rho^2}\right)\ d\phi^2 \label{rastallmetric}
\end{eqnarray}
Here, $N_s$ is the surrounding field structure parameter and 
\[
\mu=\frac{1-3\omega_s}{1-3\kappa\lambda(1+\omega_s)},\qquad
\rho^2=r^2 +a^2 \cos^{2}\theta .
\]
The horizon structure of this spacetime is determined by the positive real
roots of
\begin{equation}
\Delta(r)=r^2-2Mr+a^2+q^2-N_s r^\mu=0 .
\label{Delta_rastall}
\end{equation}
Unlike the standard KN spacetime, the presence of the term
$N_s r^\mu$ modifies the usual horizon condition $M^2\geq a^2+q^2$.
For the parameter ranges considered below, in which the positive $r$ CTC region is found to disappear, we take $N_s>0$ and the corresponding exponent
satisfies $\mu<0$. In this case,
\begin{equation}
\lim_{r\rightarrow 0^+}\Delta(r)=-\infty,
\qquad
\lim_{r\rightarrow\infty}\Delta(r)=+\infty .
\end{equation}
Hence, by continuity, $\Delta(r)=0$ always possesses at least one positive real
root. Therefore, the parameter choices considered below are not subject to the
standard KN bound $M^2\geq a^2+q^2$. 

The $g_{\phi\phi}$ component is given by:

\begin{eqnarray}
   g_{\phi\phi} &=& \sin^2\theta \left( r^2+a^2+\frac{a^2(2Mr-q^2+N_sr^{\mu})\sin^2\theta}{\rho^2}\right)\ \\
  &=&    r^2+a^2+\frac{a^2(2Mr-q^2+N_sr^{\mu})}{r^2}\ \quad (\text{On the equatorial plane }\theta =\pi/2) \label{gphiphi_rastall}
\end{eqnarray}

We note that $g_{\phi\phi}$ depends not only on the black-hole parameters
$(M,a,q)$ but also on the surrounding matter distribution through the
equation-of-state parameter $\omega_s$ and on the Rastall coupling
through the exponent $\mu$. Consequently, the chronology-violating
region is expected to depend on both the modified gravitational
coupling and the nature of the surrounding matter field.

 On the equatorial plane, let us define
\begin{equation}
\mathcal{Q}(r)\equiv r^2 g_{\phi\phi}
= r^4+a^2r^2+a^2\left(2Mr-q^2+N_s r^\mu\right)=\left(r^2+a^2\right)^2-a^2\Delta(r).\label{Qfunction}
\end{equation}
Let $r_h>0$ denote any positive root of $\Delta(r)=0$. At the horizon,
$$
N_s r_h^\mu
=
r_h^2-2Mr_h+a^2+q^2.
$$
Substituting this into Eq.~(\ref{Qfunction} gives
$
\mathcal{Q}(r_h)
>0,
$
and hence
\begin{equation}
g_{\phi\phi}(r_h)
=
\frac{\left(r_h^2+a^2\right)^2}{r_h^2}>0.
\end{equation}
Thus, independently of $q$, $N_s$, $\mu$, and the surrounding matter
distribution, a positive $r$ horizon itself cannot belong to the
equatorial CTC region, nor can the boundary $g_{\phi\phi}=0$ coincide
with a horizon. Moreover, from the expression of $\mathcal{Q}(r)$,
the condition $g_{\phi\phi}<0$ implies
\begin{equation}
\Delta(r)>
\frac{\left(r^2+a^2\right)^2}{a^2}>0.
\end{equation}
Therefore, whenever an equatorial CTC region is present, it must lie
entirely within a region where $\Delta(r)>0$ and cannot cross any
horizon.
Unlike the braneworld case discussed in Sec.~\ref{sec:Tidal}, however, $\mathcal{Q}(r)$ is
not necessarily monotonic in the Rastall geometry. Indeed,
\begin{equation}
\frac{d\mathcal{Q}}{dr}
=
4r^3+2a^2r+2Ma^2+a^2N_s\mu r^{\mu-1},
\end{equation}
whose sign is not fixed for arbitrary values of the Rastall parameters.
Consequently, the above argument establishes that the CTC region cannot
cross a horizon, but by itself does not imply a parameter-independent
ordering of the CTC boundary relative to every positive root of
$\Delta(r)$. The location of the CTC region relative to the inner
horizon must therefore be determined for the particular black-hole
branch and parameter range under consideration.

Next, we consider three representative
surrounding matter distributions,

\begin{enumerate}
\item Dust ($\omega_s=0$),
\item Dark energy ($\omega_s=-2/3$),
\item Perfect dark matter ($\omega_s=-1/3$).
\end{enumerate}

For these choices, the exponent $\mu$ reduces respectively to

\begin{equation}
\mu=
\frac{1}{1-3\kappa\lambda},
\qquad
\frac{3}{1-\kappa\lambda},
\qquad
\frac{2}{1-2\kappa\lambda},
\label{mu_definitions}
\end{equation}

which enables a direct numerical investigation of the influence of the
Rastall coupling on the causal structure.
The critical Rastall coupling at a fixed radius can  be obtained in closed form by setting $g_{\phi\phi}(r_0)=0$ in Eq.~(\ref{gphiphi_rastall}),
\begin{equation}
N_s\, r_0^{\mu} = \Phi(r_0), \qquad
\Phi(r_0) \equiv q^2 - 2Mr_0 - r_0^2 - \frac{r_0^4}{a^2},
\end{equation}
so that, whenever $\Phi(r_0)/N_s>0$,
\begin{equation}
\label{eq:mucrit}
\mu_{\mathrm{crit}}(r_0) = \frac{\ln\!\big(\Phi(r_0)/N_s\big)}{\ln r_0}.
\end{equation}
Since each relation in Eq.~(\ref{mu_definitions}) is a M\"obius function of $\kappa\lambda$, inverting it at
$\mu=\mu_{\mathrm{crit}}(r_0)$ yields the critical Rastall coupling as
\begin{align}
\text{Dust:} \qquad & \kappa\lambda_{\mathrm{crit}}(r_0) = \frac{\mu_{\mathrm{crit}}(r_0)-1}{3\,\mu_{\mathrm{crit}}(r_0)}, \\
\text{Dark energy:} \qquad & \kappa\lambda_{\mathrm{crit}}(r_0) = \frac{\mu_{\mathrm{crit}}(r_0)-3}{\mu_{\mathrm{crit}}(r_0)}, \\
\text{Perfect dark matter:} \qquad & \kappa\lambda_{\mathrm{crit}}(r_0) = \frac{\mu_{\mathrm{crit}}(r_0)-2}{2\mu_{\mathrm{crit}}(r_0)}.
\end{align}
The intervals reported in Table~1 correspond to the extrema of $\kappa\lambda_{\mathrm{crit}}(r_0)$
over the positive-$r_0$ domain for each matter model. We note that $\mu_{\mathrm{crit}}(r_0)$ in Eq.~\eqref{eq:mucrit} is real only where
$\Phi(r_0)/N_s>0$; outside this range the critical coupling must be obtained numerically.

For each surrounding matter distribution, we analyse the radial behaviour of $g_{\phi\phi}$ for various values of the Rastall coupling
$\kappa\lambda$ while keeping the remaining parameters fixed. Since the
chronology-violating region is characterised by
$g_{\phi\phi}<0$, these radial profiles identify the intervals of the
radial coordinate for which closed timelike curves are present.
\begin{figure}[htbp]
    \centering

    \begin{subfigure}{0.3\textwidth}
        \centering
        \includegraphics[width=\linewidth]{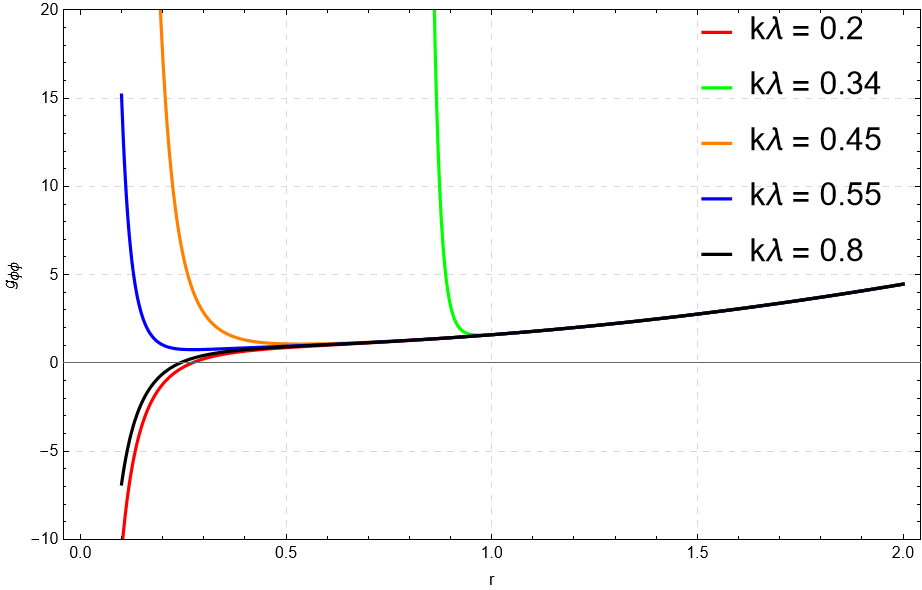}
        \caption{Dust}
    \end{subfigure}
    \hfill
    \begin{subfigure}{0.3\textwidth}
        \centering
        \includegraphics[width=\linewidth]{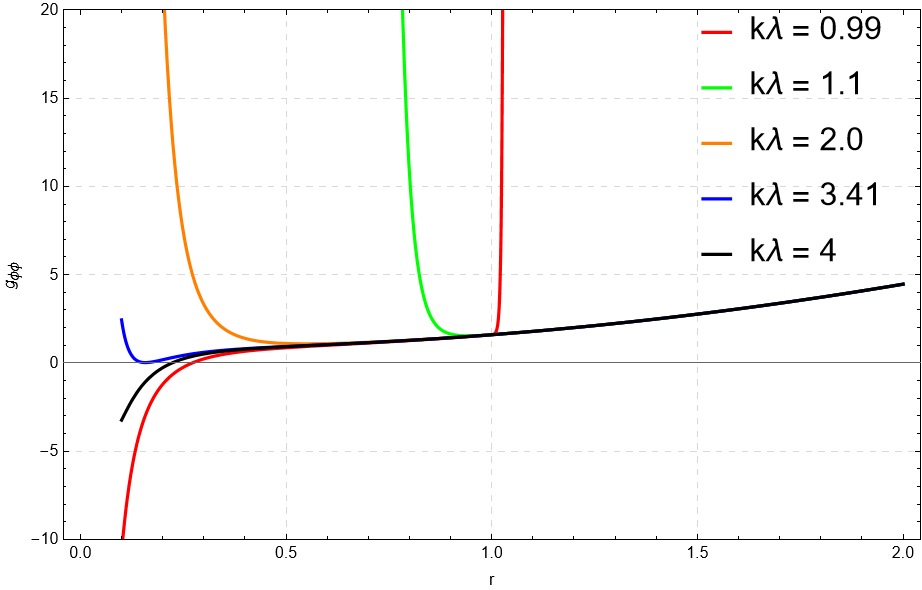}
        \caption{Dark Energy}
    \end{subfigure}
    \hfill
    \begin{subfigure}{0.3\textwidth}
        \centering
        \includegraphics[width=\linewidth]{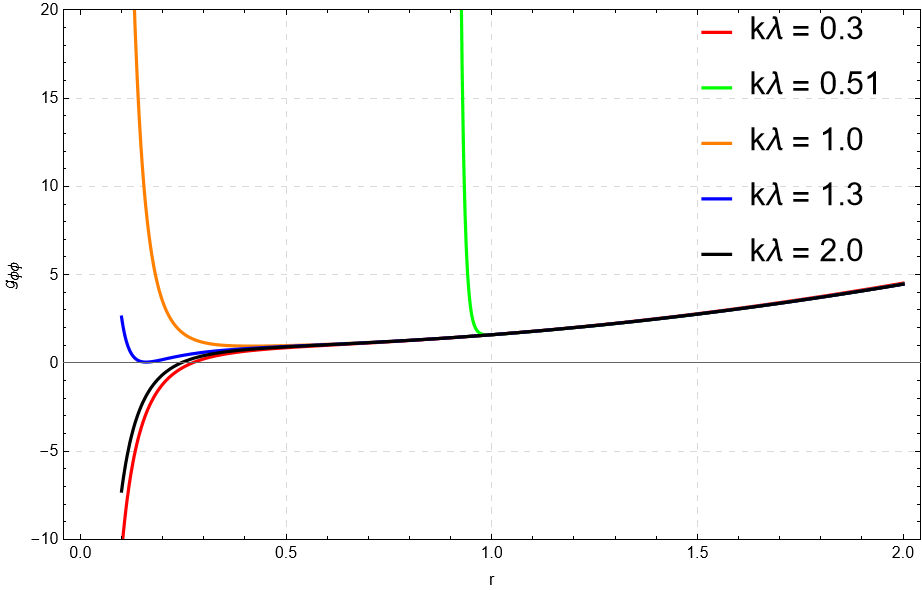}
        \caption{Dark matter}
    \end{subfigure}
    \caption{$g_{\phi\phi}$ vs $r$ plots for Kerr-Newman black hole in Rastall gravity for different values of $\kappa \lambda \qquad$ (a) Dust (b) Dark energy and (c) Dark matter.}
    \label{fig:Rastall}
\end{figure}

Fig.~\ref{fig:Rastall} shows that the Rastall coupling has a significant influence
on the chronology-violating region for all three surrounding matter
distributions. Although the detailed behaviour depends on the
equation of state, each case exhibits a finite interval of
$\kappa\lambda$ for which the positive $r$ CTC region disappears.

For the representative choice
$M=1$, $a=0.5$, $q=0.8$, and $N_s=0.03$, the corresponding parameter intervals are
summarised in Table~\ref{tab:rastall_ctc}.

\begin{table}[H]
\centering
\begin{tabular}{lc}
\toprule
\textbf{Surrounding Matter} & \textbf{Interval of $\kappa\lambda$} \\
\midrule
Dust & $\frac{1}{3} < \kappa\lambda \le 0.6$ \\
Dark Energy & $1.00 < \kappa\lambda \le 3.41$ \\
Perfect Dark Matter & $\frac{1}{2} < \kappa\lambda \le 1.3$ \\
\bottomrule
\end{tabular}
\caption{Parameter intervals of the Rastall coupling for which the
positive-$r$ equatorial CTC region disappears for different
surrounding matter distributions. The results correspond to
$M=1$, $a=0.5$, $q=0.8$, and $N_s=0.03$.}
\label{tab:rastall_ctc}
\end{table}
We now keep the radial
coordinate fixed and study the variation of $g_{\phi\phi}$ with
$\kappa\lambda$. The intersections of the curves with
$g_{\phi\phi}=0$ identify the critical values of the Rastall coupling
at which the selected point enters or leaves the chronology-violating
region.

\begin{figure}[H]
    \centering

    \begin{subfigure}[b]{0.45\textwidth}
        \centering
        \includegraphics[width=\textwidth]{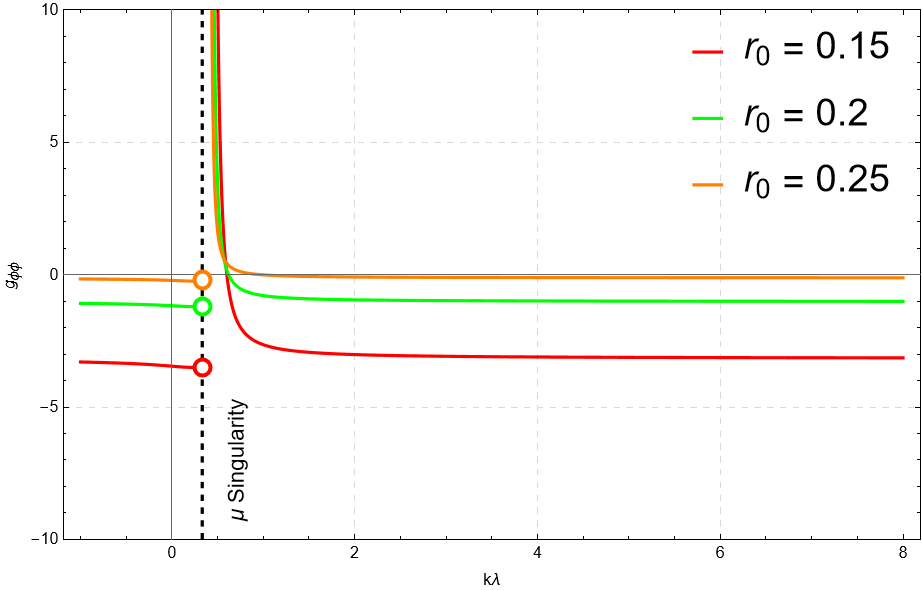}
        \caption{Dust}
    \end{subfigure}
    \hfill
    \begin{subfigure}[b]{0.45\textwidth}
        \centering
        \includegraphics[width=\textwidth]{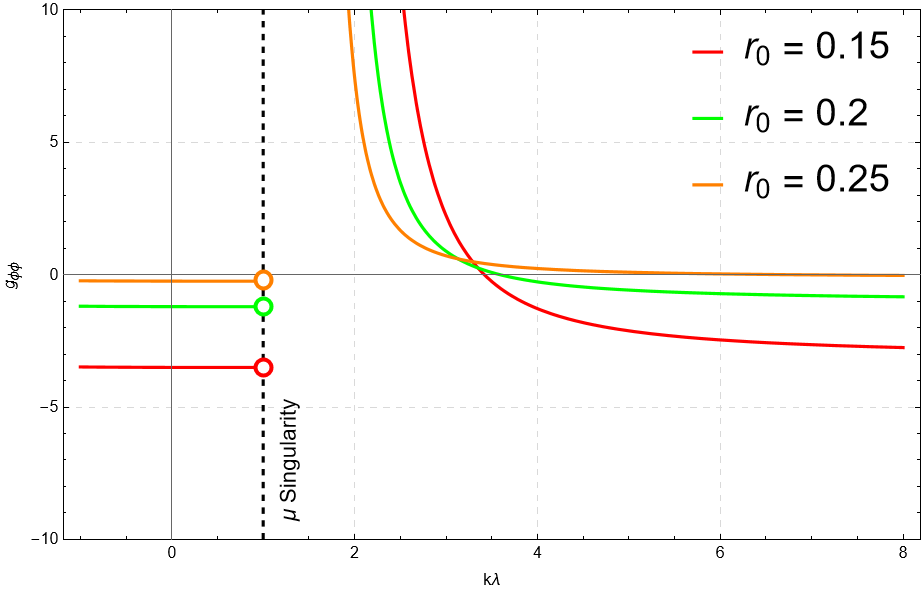}
        \caption{Dark Energy}
    \end{subfigure}

    \vspace{0.5cm}

    \begin{subfigure}[b]{0.45\textwidth}
        \centering
        \includegraphics[width=\textwidth]{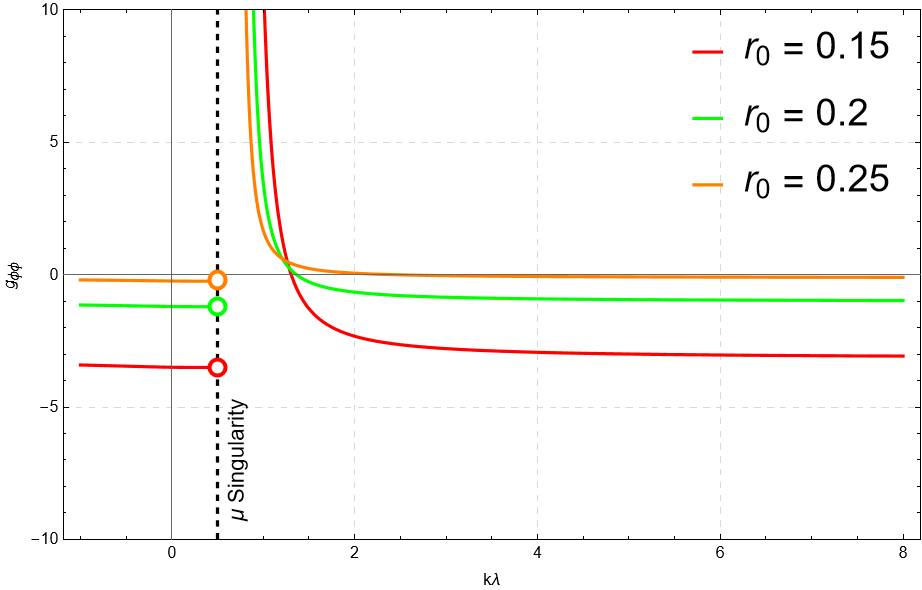}
        \caption{Dark matter}
    \end{subfigure}

    \caption{Variation of the equatorial metric component \(g_{\phi\phi}\) with the Rastall coupling \(\kappa\lambda\) at fixed radial positions for (a) dust, (b) dark energy, and (c) perfect fluid dark matter.}
    \label{fig:Rastall-gphiphi-klam}
\end{figure}
This point is illustrated in Fig.~\ref{fig:Rastall-gphiphi-klam}. We note that as $\kappa\lambda$ is varied,
$g_{\phi\phi}$ changes sign at a critical value, indicating that the
chosen spacetime point changes from being causally well behaved to
lying inside the CTC region, or vice versa. In addition to these zero crossings, the curves also exhibit singular behaviour at particular values of $\kappa\lambda$. This originates from the divergence of the exponent $\mu$ in Eq.~(\ref{mu_definitions}), which occurs at $\kappa\lambda=1/3$, $1$, and $1/2$ for dust, dark energy, and perfect dark matter, respectively. These singular values are therefore excluded from the corresponding parameter domains, and the curves on their two sides represent disconnected branches.

\subsection{Johannsen-Psaltis Kerr-Newman Spacetime}\label{sec:JPKN}

The Johannsen-Psaltis (JP) spacetime provides a phenomenological
framework for studying possible deviations from the Kerr geometry in
the strong-field regime. The motivation for introducing such a
parametrisation comes from the black-hole no-hair theorem \cite{Chrusciel:2012jk}, according
to which an isolated, stationary and asymptotically flat black hole in
general relativity is completely characterised by its mass, angular
momentum and electric charge. A parametrised deformation of the Kerr
geometry therefore provides a useful way of investigating possible
departures from the standard black-hole spacetime without specifying
the underlying field equations.

Johannsen and Psaltis~\cite{JohannsenPsaltis2011} introduced a family of
deformed Kerr metrics in which deviations from the Kerr geometry are
described by a set of dimensionless parameters $\epsilon_k$. The
deformation is introduced through the function

\begin{equation}
h(r,\theta)
=
\sum_{k=0}^{\infty}
\left(
\epsilon_{2k}
+
\epsilon_{2k+1}\frac{Mr}{\rho^2}
\right)
\left(
\frac{M^2}{\rho^2}
\right)^k ,
\end{equation}

where

\begin{equation}
\rho^2=r^2+a^2\cos^2\theta .
\end{equation}

The lowest-order deformation parameters are strongly restricted by
the requirement of asymptotic flatness and by weak-field
observational constraints. In particular, $\epsilon_0$ and
$\epsilon_1$ are set to zero, while $\epsilon_2$ is strongly
constrained by post-Newtonian observations. The leading parameter
that can therefore produce appreciable deviations in the strong-field
region is $\epsilon_3$. Retaining only this contribution gives

\begin{equation}
h(r,\theta)
=
\epsilon_3\frac{M^3r}{\rho^4}.
\end{equation}

The Kerr geometry is recovered in the limit $\epsilon_3=0$.
Consequently, $\epsilon_3$ provides a direct measure of the
deformation away from Kerr and is the parameter of primary interest
in our analysis. The JP construction may also be viewed alongside
other parametrisations of strong-field deviations from the Kerr
geometry, such as the dynamical Chern-Simons correction studied in \cite{YunesPretorius2009}, which introduces
corrections that scale as $r^{-4}$ in the slow-rotation expansion. 

The charged extension of the JP spacetime was
constructed by Rahim and Saifullah~\cite{RahimSaifullah2019}, starting
from a deformed Reissner--Nordstr\"om geometry and introducing
rotation through the Newman-Janis procedure. The resulting spacetime
contains the mass $M$, rotation parameter $a$, electric charge $q$,
and the deformation parameter $\epsilon_3$. It therefore provides a
natural setting for examining how a strong-field deformation of the
KN geometry affects the occurrence of CTCs.

The charged JP metric can be written as

\begin{align}
ds^2
={}&
-(1+h)
\left(
1-\frac{2Mr}{\rho^2}+\frac{q^2}{\rho^2}
\right)dt^2
-\frac{2a(2Mr-q^2)\sin^2\theta}{\rho^2}
(1+h)\,dt\,d\phi
\nonumber\\
&+
\frac{\rho^2(1+h)}
{\Delta+a^2h\sin^2\theta}\,dr^2
+\rho^2d\theta^2
\nonumber\\
&+
\left[
\sin^2\theta
\left(
r^2+a^2+
\frac{a^2(2Mr-q^2)\sin^2\theta}{\rho^2}
\right)
+
h\,
\frac{a^2\sin^4\theta
(\rho^2+2Mr-q^2)}
{\rho^2}
\right]d\phi^2 ,
\end{align}

where

\begin{equation}
\Delta=r^2-2Mr+a^2+q^2.
\end{equation}

For the study of CTCs, we again concentrate on the
azimuthal metric component. From the above metric,

\begin{equation}
g_{\phi\phi}
=
\sin^2\theta
\left(
r^2+a^2+
\frac{a^2(2Mr-q^2)\sin^2\theta}{\rho^2}
\right)
+
h(r,\theta)
\frac{a^2\sin^4\theta
(\rho^2+2Mr-q^2)}
{\rho^2}.
\end{equation}

On the equatorial plane, $\theta=\pi/2$, one has
$\rho^2=r^2$ and

\begin{equation}
h(r,\pi/2)
=
\frac{\epsilon_3M^3}{r^3}.
\end{equation}

The azimuthal component therefore reduces to

\begin{equation}
g_{\phi\phi}
=
r^2+a^2
+
\frac{a^2(2Mr-q^2)}{r^2}
+
\frac{\epsilon_3M^3a^2
(r^2+2Mr-q^2)}{r^5}. \label{psaltisgphiphiequator}
\end{equation}

The first three terms correspond to the usual KN 
contribution, whereas the last term represents the modification due
to the JP deformation. Thus, in addition to the
standard black-hole parameters, the causal structure now depends
explicitly on $\epsilon_3$. It is important to note that the electric charge enters
$g_{\phi\phi}$ only through $q^2$. Consequently, the equatorial
causal structure determined by $g_{\phi\phi}$ is invariant under
$q\rightarrow -q$. We therefore take $q\geq0$ in the following
analysis, with $q$ representing the magnitude of the electric charge.

Another important feature of Eq.~(\ref{psaltisgphiphiequator}) is that, at a fixed radial
position, $g_{\phi\phi}$ depends linearly on the deformation
parameter. Writing

\begin{equation}
g_{\phi\phi}(r,\epsilon_3)
=
g_{\phi\phi}^{\rm KN}(r)
+
\epsilon_3\,
\frac{M^3a^2(r^2+2Mr-q^2)}{r^5}, \label{JPgphiphi2}
\end{equation}

where

\begin{equation}
g_{\phi\phi}^{\rm KN}(r)
=
r^2+a^2+
\frac{a^2(2Mr-q^2)}{r^2},
\end{equation}

makes the role of the deformation parameter particularly clear.
For a fixed radius $r=r_0$, the critical value of $\epsilon_3$ at which the selected
point lies on the boundary of the CTC region is obtained from
$g_{\phi\phi}(r_0,\epsilon_{3, \mathrm{crit}})=0$. This gives

\begin{equation}
\epsilon_{3,\mathrm{crit}}(r_0)
=
-
\frac{
r_0^5\,g_{\phi\phi}^{\rm KN}(r_0)
}{
M^3a^2(r_0^2+2Mr_0-q^2)
},
\label{eq:eps3crit}
\end{equation}

provided that
$r_0^2+2Mr_0-q^2\neq0$.

The direction in which varying $\epsilon_3$ moves the selected point
towards or away from the CTC region is determined by

\begin{equation}
\frac{\partial g_{\phi\phi}}{\partial\epsilon_3}
=
\frac{M^3a^2(r^2+2Mr-q^2)}{r^5}. \label{e3dergphiphi}
\end{equation}
Hence the effect of increasing $\epsilon_3$ is not universal over
the entire radial domain. Its sign depends on
$r^2+2Mr-q^2$. In regions where this quantity is positive,
increasing $\epsilon_3$ increases $g_{\phi\phi}$, whereas, where it is
negative, the same change decreases $g_{\phi\phi}$. This feature is
important when interpreting the numerical plots and shows that the
effect of the deformation parameter depends on both the charge and
the radial position.

It is also useful to compare the CTC boundary with the horizon structure.
For a stationary and axisymmetric spacetime, the Killing horizon satisfies
$g_{tt}g_{\phi\phi}-g_{t\phi}^{2}=0$. For the charged
JP metric, this condition on the equatorial plane becomes
\begin{equation}
    (1+h)\left(\Delta(r)+a^2h\right)=0.
\end{equation}
The branch continuously connected to the KN horizons is
therefore determined by $\Delta(r)+a^2h=0$, which yields
\begin{equation}
   \epsilon_3^{\rm hor}(r)
=
-\frac{r^3(r^2+a^2+q^2-2Mr)}{a^2M^3}.
\end{equation}
This expression gives, for each radial position, the value of the deformation parameter for which that point lies on the Killing horizon. It can therefore be compared directly with the critical value $\epsilon_{3,\rm crit}(r)$ obtained from $g_{\phi\phi}=0$, which determines the boundary of the CTC region. Equivalently, evaluating Eq.~(\ref{JPgphiphi2}) on this Killing-horizon branch gives
\begin{equation}
    g_{\phi\phi}(r_h)
= \frac{(2Mr_h-q^2)^2}{r_h^2}\geq0.
\end{equation}
Thus, the equatorial Killing horizon itself does not belong to the CTC region. A more detailed analysis in \cite{RahimSaifullah2019} shows that, on the equatorial plane, the CTC region lies inside the inner Killing horizon and therefore does not extend outside the inner region.

The pointwise criterion Eq.~(\ref{eq:eps3crit}) can be extended to a global condition on the positive-$r$ domain. From Eq.~(\ref{e3dergphiphi}), the sign of $\partial g_{\phi\phi}/\partial\epsilon_3$ changes at
\begin{equation}
    r_*=-M+\sqrt{M^2+q^2}.
\end{equation}
For $r>r_*$, one has $r^2+2Mr-q^2>0$ and hence
$\partial g_{\phi\phi}/\partial\epsilon_3>0$, whereas for $0<r<r_*$,
we have
$\partial g_{\phi\phi}/\partial\epsilon_3<0$. Noting that $g_{\phi\phi}$ is linear in $\epsilon_3$ at fixed $r$, the requirement $g_{\phi\phi}(r,\epsilon_3)>0$ becomes
\begin{equation}
\epsilon_3>\epsilon_{3,\mathrm{crit}}(r)\qquad (r>r_*),
\qquad 
\epsilon_3<\epsilon_{3,\mathrm{crit}}(r)\qquad (0<r<r_*).
\label{eq:pointwise}
\end{equation}

The complete positive $r$ equatorial CTC region disappears when
$g_{\phi\phi}(r,\epsilon_3)>0$ for every $r>0$. Thus, the first condition in Eq.~(\ref{eq:pointwise}) requires $\epsilon_3$ to exceed the maximum value of $\epsilon_{3,\mathrm{crit}}(r)$ over $(r_*,\infty)$, while the second requires $\epsilon_3$ to lie below the minimum value of $\epsilon_{3,\mathrm{crit}}(r)$ over $(0,r_*)$. At $r=r_*$, the deformation-dependent term vanishes and $g_{\phi\phi}(r_*)=r_*^2>0$, independently of $\epsilon_3$.

We therefore define
\begin{equation}
\epsilon_3^- \equiv
\max_{r>r_*}\epsilon_{3,\mathrm{crit}}(r),
\qquad
\epsilon_3^+ \equiv
\min_{0<r<r_*}\epsilon_{3,\mathrm{crit}}(r).
\end{equation}
If these extrema occur at $r=\bar r_2\in(r_*,\infty)$ and
$r=\bar r_1\in(0,r_*)$, respectively, then
\[
\left.\frac{d\epsilon_{3,\mathrm{crit}}}{dr}\right|_{r=\bar r_1}=0,
\qquad
\left.\frac{d\epsilon_{3,\mathrm{crit}}}{dr}\right|_{r=\bar r_2}=0,
\]
with
\[
\epsilon_3^-=\epsilon_{3,\mathrm{crit}}(\bar r_2),
\qquad
\epsilon_3^+=\epsilon_{3,\mathrm{crit}}(\bar r_1).
\]
The CTC-free window is therefore
\begin{equation}
\epsilon_3^-<\epsilon_3<\epsilon_3^+ .
\label{eq:eps3window}
\end{equation}

We now examine the radial behaviour of $g_{\phi\phi}$ for
different values of the deformation parameter $\epsilon_3$, while
keeping the remaining spacetime parameters fixed. This allows us to
study how the JP deformation changes the
location and extent of the chronology-violating region.
\begin{figure}[H]
\centering
    \includegraphics[width=0.5\linewidth]{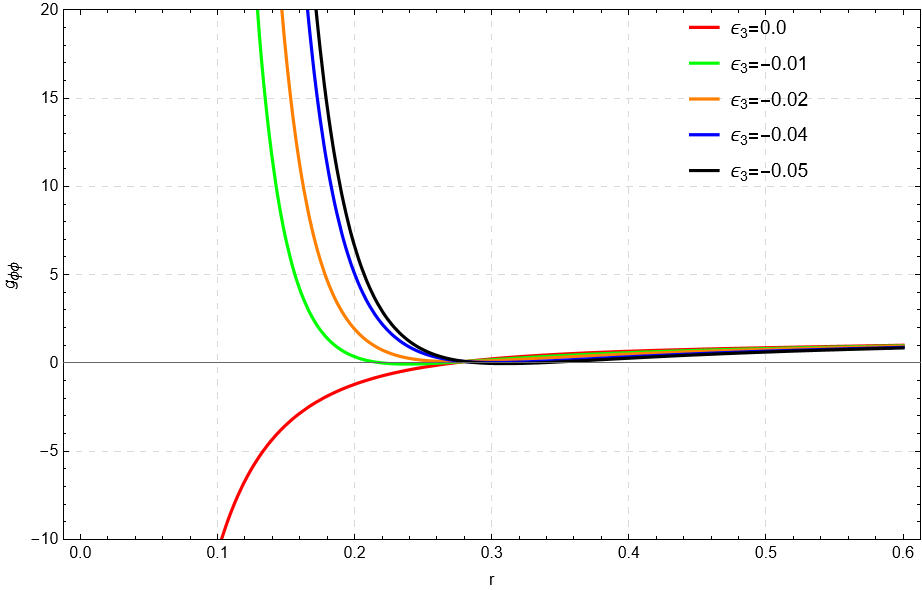}
\caption{Radial behaviour of the equatorial azimuthal metric
component $g_{\phi\phi}$ in the charged Johannsen--Psaltis spacetime
for different values of the deformation parameter $\epsilon_3$, with
$M=1$, $a=0.5$, $q=0.8$. The region satisfying
$g_{\phi\phi}<0$ corresponds to the chronology-violating region. Varying \(\epsilon_3\) modifies the extent of this region and, for suitable values of the deformation parameter, removes it from the positive radial domain shown.}
\label{fig:JP-r}
\end{figure}
Fig.~5 shows the corresponding radial profiles. For the chosen values of the parameters, the undeformed case $\epsilon_3=0$ contains a region with $g_{\phi\phi}<0$ at small positive $r$. As we vary $\epsilon_3$, the behaviour of $g_{\phi\phi}$ near the central region changes significantly and the chronology-violating region is progressively modified. For $M=1$, $a=0.5$, and $q=0.8$, the limiting values defined above are found numerically to be
$\epsilon_3^-\simeq-0.0439$ and $\epsilon_3^+\simeq-0.0124$. Thus, the CTC-free window is
\[
-0.0439<\epsilon_3<-0.0124,
\]
in agreement with Eq.~(\ref{eq:eps3window}). For values of $\epsilon_3$ within this interval, $g_{\phi\phi}$ remains positive throughout the entire positive radial domain, not only over the range displayed in Fig.~5. Thus, the deformation parameter can eliminate the positive $r$ equatorial CTC region for an appropriate range of its values.

\begin{figure}[H]
\centering
    \includegraphics[width=0.5\textwidth]{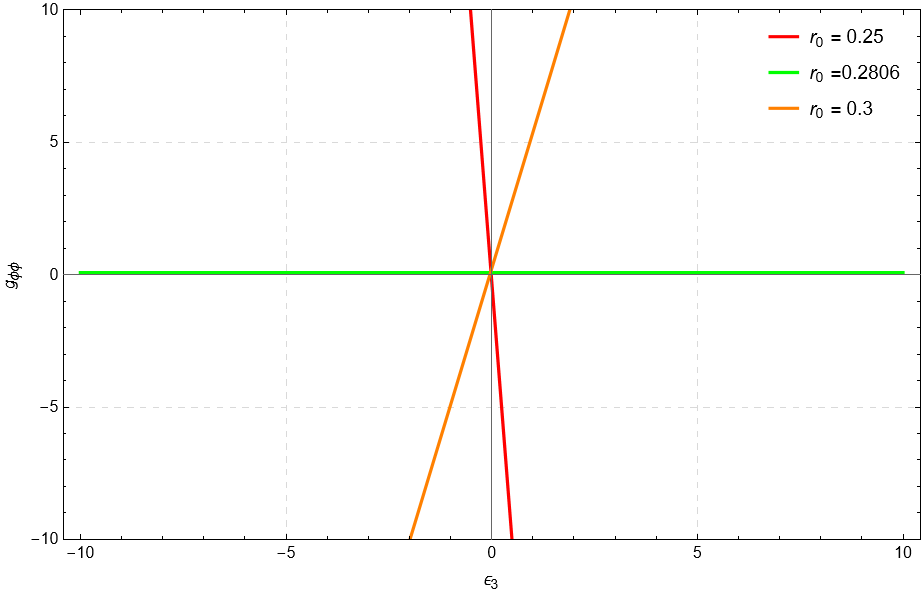}
\caption{Variation of the equatorial metric component
$g_{\phi\phi}$ with the Johannsen--Psaltis deformation parameter
$\epsilon_3$ at three fixed radial positions. The three curves
illustrate the negative, zero, and positive slopes predicted by
Eq.~(\ref{e3dergphiphi}). For $M=1$, $a=0.5$, $q=0.8$, the intermediate radial
value $r_0=\frac{\sqrt{41}}{5}-1\simeq0.2806$ satisfies
$r_0^2+2Mr_0-q^2=0$, so that $g_{\phi\phi}$ is independent of
$\epsilon_3$ at this radius. The crossings with
$g_{\phi\phi}=0$ determine the critical deformation parameters at
which the corresponding radial points enter or leave the CTC region.}
\label{fig:JP-epsilon}
\end{figure}

To further elucidate how the deformation parameter controls
the causal character of a fixed spacetime point, we keep the
radial coordinate fixed and vary $\epsilon_3$. Since
$g_{\phi\phi}$ depends linearly on $\epsilon_3$ at fixed $r=r_0$, the
resulting curves are straight lines whose slopes are determined by
Eq.~(\ref{e3dergphiphi}). Fig.~\ref{fig:JP-epsilon} shows this behaviour for three representative radial
positions. The effect of varying $\epsilon_3$ is clearly different
at different radii. Depending on the sign of
$r_0^2+2Mr_0-q^2$, the slope can be positive, negative, or vanish.
At the radius where it vanishes, $g_{\phi\phi}$ becomes independent of $\epsilon_3$.

For the other two radial positions, the intersections with $g_{\phi\phi}=0$ determine the corresponding local critical values $\epsilon_{3,\mathrm{crit}}(r_0)$. Thus, depending on the radial position, varying $\epsilon_3$ through its critical value can move the selected point either into or out of the CTC region. It should be emphasised that this fixed-radius transition represents only a local change in the causal character of the spacetime. The fact that $g_{\phi\phi}$ becomes positive at a particular value of $r$ does not by itself imply that the entire chronology-violating region has disappeared. The latter requires $g_{\phi\phi}>0$ throughout the positive radial domain, as discussed above in obtaining the window $\epsilon_3^-<\epsilon_3<\epsilon_3^+$. For the parameters used in Fig.~6, the curves at $r_0=0.25$ and $r_0=0.3$ therefore have slopes of opposite signs, in agreement with Eq.~(\ref{e3dergphiphi}).

\section{Conclusion}\label{sec:Conclusion}
In this paper, we studied the chronology protection problem and the possibility of
eliminating CTCs for three specific rotating black hole models in modified gravity: the
tidally charged braneworld, KN in Rastall gravity, and the Johannsen-Psaltis
KN spacetimes. Motivated by our previous work~\cite{Prasad:2018hdj}, we examined how the extra parameters
present in these models alter the areas containing closed timelike curves (CTCs). For the sake of simplicity, 
we restricted our analysis to the CTCs occurring on the  equatorial plane identified
through the condition $g_{\phi\phi}<0$. First, we studied the radial behaviour of
$g_{\phi\phi}$ for different values of the additional parameters to determine the extent of
the CTC region. Second, we fixed the radial coordinate and varied the relevant parameter to
identify the critical value at which a given spacetime point enters or leaves the
chronology-violating region.

As shown analytically in Sec.~\ref{sec:Tidal}, the tidal charge alone determines whether the braneworld black hole admits a positive $r$ equatorial CTC region, with 
$q \le 0$ removing it entirely. This is the only one of the three cases considered here, for which the global dependence on the control parameter is monotonic and the condition for eliminating the positive-$r$ CTC region can be obtained in a simple closed form.
In this case, the CTC region can also be shown to lie entirely inside the inner horizon,
so that it remains inaccessible to exterior observers whenever it is present.
In the case of
the KN black hole in Rastall gravity, the CTC region is controlled 
by both the Rastall coupling parameter and the surrounding matter distribution. For
the representative parameter values considered in this work, dust, dark energy and
perfect dark matter each admit finite intervals of the Rastall coupling for which
the positive $r$ equatorial CTC region is absent. This shows that
different matter environments modify the chronology violating region in
quantitatively different ways, providing an example in which additional degrees of freedom introduced by a modified theory of gravity can effectively suppress
chronology violation within the parameter range considered.
We also showed that, independently of the surrounding matter distribution, the CTC region
cannot cross any horizon of this spacetime; establishing its position relative to the inner
horizon in general, however, requires case-by-case analysis, since $g_{\phi\phi}$ need not be
monotonic in this geometry.
For the
Johannsen-Psaltis KN spacetime, the deformation parameter $\epsilon_3$ changes the
CTC region in a  different manner. The dependence of $g_{\phi\phi}$ on $\epsilon_3$ is
linear, while the sign of its variation (relative to $\epsilon_3$) is determined by the quantity
$r^2+2Mr-q^2$. Consequently, the same deformation parameter can either
increase or decrease $g_{\phi\phi}$ depending on the radial location. The special radius satisfying $r^2+2Mr-q^2=0$ separates these two behaviours and provides
an analytical explanation for the numerical results. Suitable values of $\epsilon_3$ can reduce the positive-$r$
chronology violating region and within the parameter ranges considered here,
remove it completely. The charged JP spacetime thus provides a
third example in which a parameter describing deviations from the standard
KN geometry can effectively control the occurrence and extent of CTCs. 

The present analysis is restricted to axial CTCs identified through
$g_{\phi\phi}<0$ on the equatorial plane, and therefore it should not be
viewed as a general proof of chronology protection in modified gravity.
Nevertheless, the results show that the causal structure of rotating
spacetimes is highly sensitive to the additional parameters introduced by
modified gravity theories. It would be interesting to extend this
analysis to other rotating solutions and to investigate the relation
between CTCs, horizon structure and energy conditions within a common
framework. 

Another interesting direction is to investigate the chronology-violating
regions obtained here from the viewpoint of the Raychaudhuri equation
\cite{Raychaudhuri:1953yv, Kar:2006ms}. In the present analysis, the occurrence
of axial CTCs is determined directly from the metric through the condition
$g_{\phi\phi}<0$ on the equatorial plane. The Raychaudhuri equation, on
the other hand, describes the evolution of causal congruences in terms of
their expansion, shear, vorticity and the curvature of the spacetime. It
therefore provides a  geometrical framework for examining
the local properties of the spacetime containing chronology-violating
regions. It would be useful to examine the behaviour of the
kinematical quantities entering the Raychaudhuri equation on approaching
and crossing the boundary $g_{\phi\phi}=0$. The curvature term
$R_{\mu\nu}u^\mu u^\nu$ further provides a connection with the energy
conditions, which have played an important role in discussions of
chronology protection \cite{Hawking:1991nk} and in studies of the relation between
energy-condition violations and CTCs \cite{Gutti:2022qsv}.
A systematic comparison of these quantities inside and outside the CTC
region may therefore help determine whether the appearance or suppression
of chronology violation is accompanied by characteristic changes in
geodesic focusing, vorticity, or the relevant energy conditions.
\section*{Acknowledgments}

S.K. is supported by the University Grants Commissions Faculty Recharge Programme (UGC-FRP), Govt. of India, New Delhi, India. This work is supported by the ASPIRE grant of Savitribai Phule Pune University (SPPU), Pune. A.S.R. is grateful for the hospitality of the Department of Physics, SPPU, Pune, and acknowledges its support. We also thank Mr. Suresh Sheldon for his initial participation in this work.   
\printbibliography
\end{document}